\documentclass{article}

\usepackage{amssymb}
\usepackage{amsmath}

\usepackage{graphicx}

\newtheorem{theorem}{Theorem}[section]
\newtheorem{prop}[theorem]{Proposition}
\newtheorem{lem}[theorem]{Lemma}

\newtheorem{remark}[theorem]{Remark}

\begin{document}
\title{Birkhoff strata of the Grassmannian Gr$\mathrm{^{(2)}}$: Algebraic curves}
\author{
B.G.Konopelchenko $^1$ and G.Ortenzi $^2$\\
$^1$ {\footnotesize Dipartimento di Fisica, Universit\`{a} del Salento }  \\
{\footnotesize and INFN, Sezione di Lecce, 73100 Lecce, Italy, \texttt{konopel@le.infn.it }} \\
 $^2$ {\footnotesize Dipartimento di Matematica Pura ed Applicazioni, } \\
{\footnotesize Universit\`{a} di Milano Bicocca, 20125 Milano, Italy, \texttt{giovanni.ortenzi@unimib.it} }
}
\maketitle
\begin{abstract}
Algebraic varieties and curves arising in Birkhoff strata of the Sato Grassmannian
Gr${^{(2)}}$ are studied. It is shown that the big cell $\Sigma_0$ contains the tower of families of the normal rational curves of all odd orders. Strata $\Sigma_{2n}$, $n=1,2,3,\dots$ have hyperelliptic subsets $W_{2n}$ 
with the points containing hyperelliptic curves of genus $n$ and their coordinate rings. Strata $\Sigma_{2n+1}$, $n=0,1,2,3,\dots$ contain $(2m+1,2m+3)-$plane curves for $n=2m,2m-1$ $(m \geq 2)$ and $(3,4)$ and $(3,5)$ curves in $\Sigma_3$, $\Sigma_5$ respectively. Curves in the strata $\Sigma_{2n+1}$ have zero genus.
\end{abstract}
\section{Introduction}
Grassmannian Gr${^{(2)}}$ is a very important specialization of the universal Sato Grassmannian \cite{SS}. The most known its appearance is due to the connection with the theory of the KdV equation \cite{SW,PS}.
The present paper is devoted to the study of the Grassmannian Gr${^{(2)}}$ within the framework proposed recently in \cite{KO}. The main idea of this approach is to analyze algebro-geometric structures arising in Sato Grassmannian, in our case in the Birkhoff strata of Gr${^{(2)}}$, without any \emph{a priori} reference to any integrable system. \par
Recall that Sato Grassmannian Gr can be viewed as the set of closed vector subspaces in the infinite dimensional set of all formal Laurent series with coefficients in $\mathbb{C}$ with certain special properties (see e.g. \cite{SW,PS}). Each subset W$\subset$Gr contains points possessing an algebraic basis $(w_0(z),w_1(z),w_2(z),\dots)$ where
\begin{equation} 
\label{wn}
 w_n=\sum_{k=-\infty}^n a_k^n z^k
\end{equation}
 of finite order $n$. Grassmannian Gr is a connected Banach manifold which exhibits a stratified structure \cite{SW,PS}, i.e. Gr=$\bigcup_{S} \Sigma_S$ where the stratum $\Sigma_S$ is a subset in Gr formed by elements of the form (\ref{wn}) such that possible values $n$ are given by the infinite set $S=\{s_0,s_1,s_2,\dots\}$ of integers $s_n$ with $s_0<s_1<s_2<\dots $ and $s_n = n$ for large $n$. Big cell $\Sigma_0$ corresponds to $S=\{0,1,2,\dots \}$. Other strata are associated with the sets $S$ different from $S_0$. \par
Gr$^{(2)}$ is the subset of elements $W$ of Gr obeying the condition $z^2 \cdot W \subset W$ \cite{SW,PS}. This condition imposes strong constraints on the Laurent series and on the structure of the strata. Namely, Birkhoff stratum $\Sigma_S$ in Gr$^{(2)}$ corresponds to the sets $S$ such that $S+2 \subset S$, i.e. all possible $S$ having the form \cite{SW,PS}
\begin{equation}
\label{Sm} 
S_m=\{-m,-m+2,-m+4,\dots,m,m+1,m+2,\dots \}
\end{equation}
with $m=0,1,2,\dots$. Codimension of $\Sigma_m$ is $m(m+1)/2$. One has Gr$^{(2)}=\bigcup_{m \geq 0} \Sigma_m$. \par
In this paper, using the properties  of the Birkhoff strata Gr$^{(2)}$, we show that the big cell $\Sigma_0$ contains a maximal closed subset $W_0$ which geometrically is a tower of infinite families of rational normal (Veronese) curves of all odd orders. It is demonstrated that the strata $\Sigma_{2n}$, $n=1,2,\dots$ contain subsets $W_{2n}$ closed with respect to pointwise multiplication if the coefficients of Laurent series $w_n$ obey certain associativity constraints. Geometrically the subsets $W_{2n}$ represent infinite families of coordinate rings for the hyperelliptic curves of genus $n$. Each point of the subset $W_{2n}$ contains hyperelliptic curves and its coordinate rings.  Then it is shown that the strata $\Sigma_3$ and $\Sigma_5$ contain $(3,4)$ and $(3,5)$ degenerate plane curves respectively. In the strata $\Sigma_{2m+1}, \ m \geq 2$ one has families of $(2m+1,2m+3)$ plane curves of zero genus. \par In the second part of this work \cite{KOprogr} the tangent cohomology of the subsets $W_n$ and associated integrable systems of hydrodynamical type will be studied.\par
The paper is organized as follows. The big cell is discussed in section 2. Stratum $\Sigma_1$ is considered in section 3. Closed subsets $W_2$ in the stratum $\Sigma_2$ and corresponding elliptic curves are studied in section 4. Stratum $\Sigma_3$ and associated $(3,4)$ curves are analysed in section 5. Section 6 is devoted to general strata 
$\Sigma_{2n}$, $(n=2,3,4,\dots)$. Stratum $\Sigma_5$ and the generic strata $\Sigma_{2n+1}$, $(n=3,4,\dots)$ are discussed in section 7.
\section{Big cell} 
The principal stratum $\Sigma_0$ for which $S=\{0,1,2,\dots\}$ (called also big cell) is a dense open set and it has codimension zero\cite{SW,PS}. It possesses a canonical basis $(p_0,p_1,p_2,\dots)$ where
\begin{equation}
 \label{0curr}
p_i(z)=z^i+\sum_{k\geq 1} \frac{H^i_k}{z^k}, \qquad i=0,1,2,\dots
\end{equation}
with arbitrary $H^i_k$.\par
Accordingly to the approach proposed in \cite{KO} we first look for a subset $W_0 \subset \Sigma_0$ closed with respect to multiplication. Similar to the big cell in the general Gr one has
\begin{lem}
\label{lem-S0}
 Laurent series (\ref{0curr}) at fixed $H^j_k$ obey the condition $z^2W_0 \subset W_0$ and the equations 
\begin{equation}
 \label{0alg}
p_j(z)p_k(z)=\sum_{l \geq 0} C_{jk}^lp_l(z)
\end{equation}
if and only if
\begin{equation}
 \begin{split}
  H^{2n}_i=& 0, \qquad i=1,2,3,\dots, n=0,1,2,\dots, \\
H^{2n+i}_{2i}=& 0, \qquad i=1,2,3,\dots, n=0,1,2,\dots \\
 \end{split}
\end{equation}
and
\begin{equation} 
\label{S0-H}
\begin{split}
&H^{2m+1}_{2(k+n)+1}-H^{{2(m+n)+1}}_{2k+1}-\sum_{s=0}^{n-1} H^{2m+1}_{2s+1}H^{2(n-s)-1}_{2k+1}=0,\\
&H^{2m+1}_{2(k+n)+1}+H^{2n+1}_{2(k+m)+1}+\sum_{l=0}^{k-1} H^{2m+1}_{2l+1}H^{2n+1}_{2(k-l)-1}=0.
\end{split}
\end{equation}
The constants $C_{jk}^l$ are given by
\begin{equation}
 \begin{split}
 C_{2n,2m}^{2l}&=\delta_{m+n}^l, \\
  C_{2n,2m+1}^{2l+1}&=\delta_{{m+n}}^l+ H^{2m+1}_{2(n-l)-1}, \\
C_{2n+1, 2m+1}^{2l}&= \delta_{{m+n+1}}^l+H^{2n+1}_{2(m-l)+1}+H^{2m+1}_{2(n-l)+1}
 \end{split}
\end{equation}
and $p_{2n}={p_2}^2=z^{2n}$, $n \geq 0$.
\end{lem}
An immediate  consequence of this lemma is given by the following
\begin{prop}
 The  subset $W_0 \subset \Sigma_0$ the elements of which are given by vector spaces with basis  $\langle p_i(z) \rangle_i$ and parameters $H^i_k$ obeying the constraints (\ref{S0-H}), is closed with respect to pointwise multiplication $W_0 \cdot W_0 \subset W_0$. It is a maximal closed subset in the big cell. This subset $W_0$ is an infinite family of infinite-dimensional associative commutative algebra with unity $p_0=1$.  
\end{prop}
The last statement follows from the equivalence of equations  (\ref{S0-H}) to the associativity conditions
\begin{equation}
 \sum_s C^s_{ij}C^r_{ks}-C^s_{ik}C^r_{js} =0
\end{equation}
for the structure constants $C_{jk}^l$. \par
 Relations (\ref{0alg}) written explicitly, i.e.
\begin{equation}
 \begin{split}
  p_{2n} p_{2m}&=p_{{2(m+n)}}, \\
  p_{2n} p_{2m+1}&=p_{{2(m+n)+1}}+\sum_{s=0}^{n-1} H^{2m+1}_{2s+1}p_{2(n-s)-1}, \\
p_{2n+1} p_{2m+1}&= p_{{2(m+n+1)}}+\sum_{s=0}^{m} H^{2n+1}_{2s+1}p_{2(m-s)}+\sum_{s=0}^{n} H^{2m+1}_{2s+1}p_{2(n-s)},
 \end{split}
\end{equation}
imply that
\begin{equation}
\label{parVer}
\begin{split} 
z^2=&p_1^2-2H^1_1, \\
p_3=&{p_1}^3-3H^1_1p_1, \\
p_5=&{p_1}^5-5H^1_1{p_1}^3+\frac{15}{2}{H^1_1}^2p_1,\\
\dots&
\end{split}
\end{equation}
 or equivalently
\begin{equation}
\label{lpi}
 \lambda={p_1}^2-2H^1_1, \qquad p_{2n+1}=\alpha_n(\lambda)p_1
\end{equation}
where $\lambda=z^2$ and $\alpha_n(\lambda)=\prod_{s=1}^n \left( \lambda-\frac{H^1_{2(n-s)+1}}{2(n-s)+1}\right)$. \par
Similar to \cite{KO} one can treat $\lambda,p_1,p_3,\dots$ as the affine coordinates. So one has the following geometrical interpretation of the subset $W_0$.
\begin{prop}
\label{prop-S0}
 Big cell $\Sigma_0$ contains an infinite--dimensional algebraic variety $\Gamma_0$ with the ideal 
\begin{equation}
 \langle \lambda-p_1^2+2H^1_1,l^{(2)}_1,l^{(2)}_1,\dots \rangle
\end{equation}
where $l^{(2)}_n=p_{2n+1}-\alpha(\lambda) p_1$ and the variables $H^j_k$ obey the constraints (\ref{S0-H}). This variety $\Gamma_0$ is an infinite tower of infinite families of rational normal (Veronese) curves of all odd orders. 
\end{prop}
Formulas (\ref{parVer}) represent a canonical parameterization of rational normal curves (see e.g. \cite{Har}). For instance, the curves defines by the first two equations (\ref{parVer}) is the classical twisted cubic in the three-dimensional space with the coordinates $(\lambda,p_1,p_3)$. \par
There is an infinite set of independent variables among all $H^j_k$ constrained by conditions (\ref{S0-H}). A natural set of independent $H^j_k$ is given by $H^1_1,H^1_3,H^1_5,\dots$. \par 
It is also easy to see using (\ref{lpi})that the ideal $I^{(2)}_0$ contains singular ``hyperelliptic'' curves of genus zero given by the equations
\begin{equation}
 p_{2n+1}^2=(\lambda+2H^1_1) \alpha_n(\lambda)^2.
\end{equation}
Infinite family of algebraic varieties described in Proposition \ref{prop-S0} is in its turn the algebraic variety in the affine space with coordinates $p_i$, $(i=1,2,3,\dots)$ and $H^j_k$, $(j,k=1,2,3,\dots)$ defined by the quadrics
\begin{equation}
 f_{jk}=p_jp_k-p_{j+k}-\sum_{s=0}^k H^j_s p_s -\sum_{s=0}^j H^k_s p_s =0
\end{equation}
and equations (\ref{S0-H}). \par
We emphasize that an infinite tower of normal rational curves for fixed $H^j_k$ is in correspondence with a point of the subset $W_0$.
\section{Stratum $\mathbf{\Sigma}_1$}
The stratum $\Sigma_1$ is the lowest stratum different from the big cell and it corresponds to $m=1$ and $S=\{-1,1,2,\dots\}$. Due to the absence of zero order element $w_0$ the canonical basis is of the form
\begin{equation}
\label{S1-p}
 p_i(z)=z^i+H^i_0+\sum_{k \geq 1 } \frac{H^i_k}{z^k}, \qquad i=1,2,3,\dots\ .
\end{equation}
Since $(p_{-1})^2 \notin \langle p_i \rangle_{i=-1,1,2\dots} $ one should consider only $p_j$ with $j=1,2,3,\dots$.  
\begin{lem}
\label{lem-S1}
A set $W_1$ of Laurent series (\ref{S1-p}) obey the condition $z^2 \cdot W_1 \subset W_1$ and the equations
\begin{equation}
\label{S1-alg}
 p_j(z)p_k(z)=\sum_{l \geq 1} C_{jk}^l p_l(z), \qquad j,k=1,2,3,\dots 
\end{equation}
if and only if the parameters $H^j_k$ satisfy the constraints 
\begin{equation}
\label{S1-H} 
\begin{split}
&H^{2{j+k}}_0+H^{2{j}}_0H^{2{k}}_0 =0, \\
&H^{2(k+j)+1}_0+H^{2k+1}_0H^{2j+1}_0+\sum_{l=0}^{j-1}H^{2k+1}_{2l+1}H^{2(j-l)-1}_0=0,\\
&H^{2k+1}_{2(j+l)+1} + H^{2j}_0H^{2k+1}_{2l+1} - H^{2(k+j)+1}_{2l+1} - H^{2j}_0H^{2k+1}_{2l+1}-H^{2k+1}_{2(l+j)+1} - \sum_{s=0}^{j-1}H^{2k+1}_{2s+1}H^{2(j-s)-1}_{2l+1}=0, \\
&H^{2(k+j+1)}_0+H^{2k+1}_0H^{2j+1}_0+\sum_{s=0}^{k-1}H^{2(k-s)}_0H^{2j+1}_{2s+1}
+\sum_{s=0}^{j-1}H^{2(j-s)}_0H^{2k+1}_{2s+1}=0,\\
&H^{2j+1}_{2(l+k)+1}+H^{2k+1}_{2(l+j)+1}+\sum_{s=0}^{l-1}H^{2j+1}_{2s+1}H^{2k+1}_{2(l-s)-1}-
\sum_{s=0}^{k-1}H^{2j+1}_{2(l+k)+1}-\sum_{s=0}^{j-1}H^{2k+1}_{2(l+j)+1}=0
\end{split}
\end{equation}
and 
\begin{equation}
\label{S1-C} 
\begin{split} 
 C^{2l}_{2j,2k}=& \delta^l_{j+k}+H^{2j}_0 \delta^l_k+H^{2k}_0 \delta^l_j, \\
 C^{2l+1}_{2j,2k+1}=& \delta^l_{j+k}+H^{2j}_0 \delta^l_k +H^{2k}_0 \delta^l_j+ H^{2k+1}_{2(j-l)+1}, \\
 C^{2l+1}_{2j+1,2k+1}=& H^{2k+1}_0 \delta^l_j+H^{2j+1}_0 \delta^l_k, \\
C^{2l}_{2j+1,2k+1}=& \delta^l_{j+k+1}+H^{2j+1}_{2(l+k)+1}+H^{2k+1}_{2(l+j)+1}\ .
\end{split}
\end{equation}
\end{lem}
The analysis of the constraints (\ref{S1-H}) gives
\begin{equation}
 H^{2n}_i=0, \qquad n,i=1,2,3,\dots
\end{equation}
and
\begin{equation}
 H^{2n}_0=-(-H^2_0)^n, \qquad n=1,2,3,\dots,
\end{equation}
i.e.
\begin{equation}
\label{S1-p2n}
 p_{2n}(z)=z^{2n}-(-H^2_0)^n, \qquad n=1,2,3,\dots.
\end{equation}
For the elements $p_{2n+1}$ one instead has
\begin{equation}
\label{S1-pnp1} 
\begin{split}
 p_2=&{p_1}^2-2H^1_0p_1, \\
 p_3=&{p_1}^3-3H^1_0{p_1}^2-(3H^1_1-3{H^1_0}^2)p_1, \\
\dots\ . &
\end{split}
\end{equation}
Similar to the big cell one has a subset $W_1$ in $\Sigma_1$ closed with respect to multiplication which algebraically is an infinite-dimensional commutative associative algebra $A_1$ with the structure constants given by 
(\ref{S1-C}) in the basis (\ref{S1-p}). Geometrically $W_1$ is an infinite tower of families of rational normal curves of all odd orders passing through the origin $p_1=p_2=p_3=\dots=0$.\par
The fact that for the stratum $\Sigma_1$ one has results which are similar to those for big cell  is not that surprising. Indeed, taking into account the relations (\ref{S1-H}), namely 
\begin{equation}
2H^1_1-H^2_0+{H^1_0}^2=0, \qquad H^3_0+H^1_0H^2_0+H^1_0H^1_1=0 
\end{equation}
and the formula (\ref{S1-p2n}), i.e. $p_2=z^2+H^2_0$, one can rewrite equations (\ref{S1-pnp1}) as 
\begin{equation}
\label{S1-p2p3}
 \begin{split}
  z^2=&(p_1-H^1_0)^2-2H^1_1, \\
p_3-H^3_0=&(p_1-H^1_0)^3-3H^1_1(p_1-H^1_0).
 \end{split}
\end{equation}
In the variables 
\begin{equation}
\tilde{p}_1=p_1-H^1_0, \qquad \tilde{p}_3=p_3-H^3_0 
\end{equation}
the equations (\ref{S1-p2p3}) the first two equations (\ref{lpi}) for the big cell. It is a direct check that
in the variables  
\begin{equation}
 \label{S0S1}
\tilde{p}_k=p_k-H^k_0, \qquad k=1,2,3,\dots
\end{equation}
all equations (\ref{S1-pnp1}) coincide with equations (\ref{lpi}) for the big cell. \par
Thus the result for the stratum $\Sigma_1$ and big cell are connected by a simple change of variables (\ref{S0S1}). Similar situation take place for other strata $\Sigma_m$ with odd $m$.
\section{Stratum $\mathbf{\Sigma_2}$ and elliptic curves}
For the stratum $\Sigma_2$ with $S=\{-2,0,2,3,4,\dots\}$ the positive order elements of the canonical basis are given by 
\begin{equation}
\label{series-S2}
 \begin{split}
  p_0=&1+\sum_{k \geq 1} \frac{H^0_k}{z^k}, \\
p_j=&z^j+H^j_{-1}z +\sum_{k \geq 1} \frac{H^j_k}{z^k}, \qquad k, j=2,3,4,\dots\ .
\end{split}
\end{equation}
First we note that $(p_{-2})^2 \notin \langle p_i \rangle_{i=-2,0,2,3,\dots}$ and the
analogue of the Lemmas \ref{lem-S0} and \ref{lem-S1} is given by
\begin{lem}
 \label{lem-S2}
A set $W_2$ of Laurent series (\ref{series-S2}) obey the equations
\begin{equation}
\label{alg-S2}
 p_j(z) p_k(z) = \sum_{l=0,2,3,\dots} C^l_{jk}p_l(z)
\end{equation}
and the condition $z^2 W_2 \subset W_2$ is satisfied if and only if
\begin{equation}
 \label{Heven-S2}
\begin{split} 
 H^{2n}_k =&0, \qquad k=-1,1,2,3,\dots, \quad n=0,1,2,\dots, \\
 H^{2n+1}_k =&0, \qquad n,k=1,2,3,\dots 
\end{split}
\end{equation}
and 
\begin{equation} 
\label{Hodd-S2}
 \begin{split}
 &H^{2m+1}_{2(k+n)+1}-H^{2(m+n)+1}_{2k+1}-\sum_{s=-1}^{n-2}H^{2m+1}_{2s+1}H^{2(n-s)+1}_{2k+1}=0, \\
&H^{2m+1}_{2(n+k)+1}+H^{2n+1}_{2(m+k)+1}+\sum_{s=-1}^kH^{2m+1}_{2s+1}H^{2n+1}_{2(l-s)-1}=0.
 \end{split}
\end{equation}
Constants $C^l_{jk}$ are given by
\begin{equation}
\label{C-S2}
\begin{split}
C^{2l}_{2n,2m}=&  \delta_{m+n}^l,\\
C_{2n,2m+1}^{2l+1}=&   \delta^l_{m+n}+H^{2m+1}_{2(n-l)-1}, \\
C^{2l}_{2n+1,2m+1}=& \delta^l_{m+n+1}+H^{2n+1}_{2(m-l)+1}+H^{2m+1}_{2(n-l)+1} 
+H^{2n+1}_{-1}H^{2m+1}_{-1}\delta^l_2\\ &+\left(H^{2n+1}_{-1}H^{2m+1}_{1}+H^{2n+1}_{1}H^{2m+1}_{-1} \right)\delta^l_0.
\end{split}
\end{equation}
which imply
\begin{equation}
\label{alg-S2-exp}
\begin{split}
p_{2n}p_{2m}=&     p_{2(m+n)},\\
p_{2n}p_{2m+1}=&   p_{2(m+n)+1}+\sum_{k=-1}^{n-2}H^{2m+1}_{2k+1} p_{2(n-k)-1}, \\
p_{2n+1}p_{2m+1}=& p_{2(m+n+1)}+\sum_{k=-1}^{m}H^{2n+1}_{2k+1} p_{2(m-k)} +\sum_{k=-1}^{n}H^{2m+1}_{2k+1} p_{2(n-k)}\\
&+H^{2n+1}_{-1}H^{2m+1}_{-1}p_2+\left(H^{2n+1}_{-1}H^{2m+1}_{1}+H^{2n+1}_{1}H^{2m+1}_{-1} \right)
\end{split}
\end{equation}
and $p_{2n}=z^{2n}$, $n \geq 0$.
\end{lem}
As a consequence, one has
\begin{prop}
 \label{prop-S2}
The stratum $\Sigma_2$ contains a maximal closed subset $W_2$ whose elements are vector spaces with basis  
(\ref{series-S2}) and parameters $H^j_k$ obeying the constraints (\ref{Heven-S2}), (\ref{Hodd-S2}) and such that $z^2 W_2 \subset W_2$. 
\end{prop}
The relations (\ref{alg-S2}) readily imply that all $p_i(z)$ are generated by two elements $z^2$ and $p_3$.\par
Using (\ref{alg-S2-exp}) one can show that the set of independent relations (\ref{alg-S2}) is given by 
\begin{equation}
 \label{Ell-S2}
{p_3}^2=\lambda^3+2H^3_{-1}\lambda^2+\left({H^3_{-1}}^2+2H^3_1\right)\lambda+2H^3_{-1}H^3_1+2H^3_3
\end{equation}
and
\begin{equation}
 \label{oddseries-S2}
 p_{2n+1}=\left(\lambda^{n-1}-\sum_{i=0}^{n-2} H_{-1}^{2(n-i)-1} \lambda^i \right) p_3
\end{equation}
This relation is obtained obtained using iteratively the formula
\begin{equation}
 p_{2n+1}=\lambda p_{2n-1}+H^{2n-1}_{-1}p_3.
\end{equation}
\begin{prop}
 Subset $W_2$ is an infinite family of infinite-dimensional commutative associative algebra with the basis $1$, $p_2$, $p_3$, $p_4,\dots$ isomorphic to $\mathbb{C}[\lambda,p_3]/ C_6$
\end{prop}
where
\begin{equation}
 \label{C6-S2}
C_6={p_3}^2-\lambda^3-2H^3_{-1}\lambda^2-\left({H^3_{-1}}^2+2H^3_1\right)\lambda-\left(2H^3_{-1}H^3_1+2H^3_3\right).
\end{equation}
{\bf Proof} Associativity follows from the fact that the conditions (\ref{Heven-S2}) and (\ref{Hodd-S2}) are equivalent to the condition
\begin{equation}
 \sum_{s=0,2,3,\dots} C_{jk}^sC_{sl}^r = \sum_{s=0,2,3,\dots} C_{lk}^sC_{sj}^r \qquad j,k,l,r=0,2,3,\dots
\end{equation}
for the constants $C_{jk}^l$ given by (\ref{alg-S2-exp}). $\square$ \par
Treating now $\lambda,p_3,p_5$ and $H^j_k$ as affine coordinates one has the following geometrical interpretation of the subset $W_2$.
\begin{prop}
 The subset $W_2$ is an infinite dimensional algebraic variety $\Gamma_2$ in the affine space with coordinates 
$p_j$, $(j=2,3,4,\dots)$, $H^j_k$, $(j=3,5,7,\dots,\ k=-1,1,3,5,\dots )$ defined by the intersection of quadrics
\begin{equation}
 \label{fjk-S2}
f_{jk}=p_jp_k-p_{j+k}-\sum_{l=0,2,3,\dots}^{j+k-1} C^l_{jk}p_l(z)=0
\end{equation}
and quadrics (\ref{Hodd-S2}). An ideal $I^{(2)}$ of this variety is 
\begin{equation}
 \label{I2-S2}
I^{(2)}=\langle C_6, l^{(2)}_5, l^{(2)}_7, l^{(2)}_9, \dots \rangle
\end{equation}
where $l^{(2)}_{2n+1}=p_{2n+1}-\left(\lambda^{n-1}-\sum_{i=0}^{n-2} H_{-1}^{2(n-i)-1} \lambda^i \right) p_3$. 
\end{prop}
Since $W_2  \sim \mathbb{C}[\lambda,p_3]/ C_6$ one can view $\Gamma_2$ as the infinite family of coordinate rings of the elliptic curve $C_6=0$ parameterized by the variables $H^j_k$ obeying the conditions (\ref{Heven-S2}) and (\ref{Hodd-S2}). Analyzing these conditions one concludes that there is an infinite set of independent variables among all $H^j_k$, for example $H^3_{-1},H^3_{1},H^3_{3},H^3_{5},\dots$. \par
It is a direct check that the curve $C_6=0$ has genus one. So the stratum $\Sigma_2$ contains an infinite family of elliptic curves parameterized by $H^3_{-1},H^3_{1},H^3_{3}$. \par
We emphasize that each of these elliptic curves belong to a point of the subset $W_2$. So, following \cite{KO}, such points of $\Sigma_2$will be called {\it elliptic points} and the whole subset $W_2$ an {\it elliptic subset}.\par
The ideal $I^{(2)}$ contains singular hyperelliptic curves of all orders and of genus $1$ given by
\begin{equation}
 \label{Hcurves-g1-S2}
p_{2n+1}^2=\left(\lambda^{n-1}-\sum_{i=0}^{n-2} H_{-1}^{2(n-i)-1} \lambda^i \right)^2
\left(\lambda^3+2H^3_{-1}\lambda^2+\left({H^3_{-1}}^2+2H^3_1\right)\lambda+2H^3_{-1}H^3_1+2H^3_3\right)
\end{equation}
\section{Stratum $\mathbf{\Sigma_3} $: (3,4) curves of zero genus}
 Next case corresponds to $S=\{-3,-1,1,3,4,5,\dots \}$. Due to the absence of elements of orders zero and two positive elements of the canonical basis are given by
\begin{equation}
 \label{series-S3}
\begin{split}
 p_1=&z+H^j_0+\sum_{k \geq 1} \frac{H^1_k}{z^k}, \\
p_j=&z^j+H^j_{-2}z^2 + H^j_0+\sum_{k \geq 1} \frac{H^j_k}{z^k}, \qquad j=3,4,5,\dots\ .
\end{split}
\end{equation}
Since ${p_1}^2$ has order two a closed subspace can be generated  only by the elements $p_3,p_4,p_5,\dots$.
\begin{lem}
 \label{lem-S3}
A set $W_3$ of Laurent series $p_j(z)$, $j=3,4,5,\dots$ obey the equations
\begin{equation}
\label{alg-S3}
 p_j(z)p_k(z)=\sum_{l=3,4,5,\dots}C_{jk}^l p_l(z),\qquad j,k=3,4,5,\dots
\end{equation}
and the condition $z^2W_3 \subset W_3$ if and only if
\begin{equation}
\label{p-S3}
  p_j=z^j+H^j_{-2}z^2+H^j_0, \qquad j \geq 5,
\end{equation}
\begin{equation}
\label{pH-S3}
\begin{split}
&H^j_{-2}+H^{j-2}_{-2}H^{4}_{-2}-H^{j-2}_{0}=0,\\
&H^j_0+H^{j-2}_{-2}H^{4}_{0}=0 
\end{split}
\end{equation} 
and
\begin{equation}
\label{nlin-S3}
\begin{split}
&H^4_{{0}}+2\,H^3_{{0}}H^3_{{-2}}-{H^3_{{-2}}}^{2}H^4_{{-2}}-{H^4_{{-2}}}^{2}=0, \\
&{H^3_{{0}}}^{2}-{H^3_{{-2}}}^{2}H^4_{{0}}-H^4_{{-2}}H^4_{{0}}=0.
\end{split}
\end{equation} 
\end{lem}
{\bf Proof} Let us begin with the condition $z^{2n} W_3 \subset W_3$. One has 
\begin{equation}
\label{evenconst-S3}
 z^{2n}p_m(z)= z^{2n+m}+ \dots+ H^{m}_{2n-1}z+\dots\ .
\end{equation}
In $W_3$ there is no element which contains the term of order one. Hence, with necessity $H^m_{2n-1}=0$ for all $n=1,2,3,\dots$ and $n=3,4,5,\dots$, i.e.
\begin{equation}
 p_j(z)=z^j+H^j_{-2}z^2+H^j_{0}+\sum_{n \geq 1}\frac{H^j_{2n}}{z^{2n}}, \qquad j=3,4,5,\dots\ .
\end{equation}
 Then considering the product $p_{2k+1}p_{j}$ one has
\begin{equation}
\label{oddconst-S3}
 p_{2k+1}(z)p_j(z)=z^{2k+j+1}+\dots+H^j_{2k}z+\dots \ .
\end{equation}
The terms  of the order $z^i$, $i \geq 3$ can be represented as a superposition of $p_3,p_4,\dots, p_{2k+j+1}$ giving the constants $C_{jk}^l$ while the coefficient in front of $z$ should vanish. Hence $H^j_{2k}=0$ for all $k =1,2,3,\dots$. So
\begin{equation}
 p_j=z^j+H^j_{-2}z^2+H^j_0 \qquad j \geq 3\ .
\end{equation}
 The coefficients $H^j_{-2}$ and $H^j_0$ are not all independent. Indeed, the relations
\begin{equation}
\label{rel-S3}
 \begin{split}
  z^2p_3=& p_{5}+H^3_{-2}p_4,  \\
  z^2p_4=& p_{6}+H^4_{-2}p_4,  \\
  z^4p_3=& p_{7}+H^3_{-2}p_6+H^3_{0}p_4,  \\
\dots&
 \end{split}
\end{equation}
imply
\begin{equation}
\label{fH-S3}
 \begin{split}
H^5_{-2} -H^3_0 +H^3_{-2} H^4_{-2} = &0, \\
H^5_{0}  +H^3_{-2} H^4_{0} = &0, \\
H^6_{-2} -H^4_0 +{H^4_{-2}}^2= &0, \\
H^6_{0}  +H^4_{-2} H^4_{0}= &0, \\
H^7_{-2} +H^3_{-2}H^6_{-2} +H^3_{0}H^4_{-2}= &0, \\
H^7_{0}  +H^3_{-2}H^6_{0} +H^3_{0}H^4_{0}= &0, \\
\dots&
 \end{split}
\end{equation}
and so on. The relations (\ref{fH-S3}) are the lowest members of the relations (\ref{pH-S3}). Using these relations, one can express all $H^j_{-2}$, $H^j_{0}$ with $j=5,6,7,\dots$ in terms of $H^3_{-2}$, $H^3_{0}$ and $H^4_{-2}$, $H^4_{0}$. \par
Furthermore, the vanishing of the coefficients in front of $z^2$ and $z^0$ in the relation
\begin{equation}
 {p_3}^2-\left(p_6 +2 H^3_{-2}p_5+{H^3_{-2}}^2p_4+2H^3_{0}p_3\right)=0
\end{equation}
is equivalent to the conditions
\begin{equation}
 \begin{split}
  &H^6_{-2}-2\,H^3_{{-2}}H^5_{{-2}}-{H^3_{{-2}}}^{2}H^4_{{-2}}=0, \\
&H^6_0-{H^3_{{0}}}^{2}-2\,H^3_{{-2}}H^5_{{0}}-{H^3_{{-2}}}^{2}H^4_{{0}}=0.
 \end{split}
\end{equation}
Finally taking into account (\ref{fH-S3}), one gets the constraints (\ref{nlin-S3}). So there are only two independent parameters among all coefficients $H^j_{-2}$ and $H^j_0$. The simplest choice is to take $H^3_{-2}$ and $H^4_{-2}$ as independent variables. At last, the direct calculation gives
\begin{equation}
\label{C-S3} 
C_{jk}^l=\delta_{j+k}^l+H^k_{-2} \delta_{j+2}^l +H^k_{0} \delta_{j}^l+
H^j_{-2} \delta_{k+2}^l +H^j_{0} \delta_{k}^l+ H^j_{-2}H^k_{-2}\delta_4^l.  \qquad \square
\end{equation}
 An immediate consequence of the Lemma \ref{lem-S3} is given by
\begin{prop}
 \label{prop-S3}
The stratum $\Sigma_3$ contains the subset $W_3$ closed with respect to  pointwise multiplication $W_3 \cdot W_3 \subset W_3$. Elements of $W_3$ are vector spaces with basis $\langle p_i \rangle_i $ of the form (\ref{p-S3}) with $H^j_{-2},H^j_0$ obeying the constraints (\ref{pH-S3}) and (\ref{nlin-S3}). The subset $W_3$ is an infinite family of infinite-dimensional associative and commutative algebra $A_3$ with the basis (\ref{p-S3}) and structure constants (\ref{C-S3}). 
\end{prop}
A geometrical interpretation of $W_3$ is provided  by
\begin{prop}
 \label{prop2-S3}
The subset $W_3$ can be viewed as the two parametric family of algebraic varieties defined by the relations 
\begin{equation}
 \label{quadp-S3}
p_jp_k-\sum_lC^l_{jk}p_l=p_jp_k-\left(p_{j+k}+H^k_{-2} p_{j+2} +H^k_{0} p_{j}+
H^j_{-2} p_{k+2} +H^j_{0} p_{k}+ H^j_{-2}H^k_{-2}p_4 \right)=0.
\end{equation}
The ideal of this family contains the plane $(3,4)$ curve (in the terminology of \cite{BEL}) defined by the equation
\begin{equation}
 \label{(3,4)-S3}
\begin{split}
 & {p_{{4}}}^{3}-{p_{{3}}}^{4}+4\,H^3_{{-2}}p_{{3}}{p_{{4}}}^{2}-
 \left( 3\,H^4_{{-2}}-2\,{H^3_{{-2}}}^{2} \right) {p_{{3}}}^
{2}p_{{4}}- \left( -4\,H^3_{{0}}+2\,H^3_{{-2}}H^4_{{-2}
} \right) {p_{{3}}}^{3}\\ 
&- \left( 3\,H^4_{{0}}+4\,H^3_{{0}}H^3_{{-2}}+{H^3_{{-2}}}^{4}+{H^3_{{-2}}}^{2}H^4_{{-
2}} \right) {p_{{4}}}^{2}- \left( 4\,{H^3_{{-2}}}^{2}H^3_{{0
}}+8\,H^3_{{-2}}H^4_{{0}}-2\,H^3_{{-2}}{H^4_{{-2}}
}^{2}\right. \\  
& \left. -6\,H^3_{{0}}H^4_{{-2}}-2\,{H^3_{{-2}}}^{3}
H^4_{{-2}} \right) p_{{3}}p_{{4}}
- \left( -3\,H^4_{{0}}H^4_{{-
2}}+6\,{H^3_{{0}}}^{2}-6\,H^3_{{0}}H^3_{{-2}}H^4_{
{-2}}+2\,{H^3_{{-2}}}^{2}H^4_{{0}}\right. \\ 
& \left.+{H^3_{{-2}}}^{2}{H^4_{{-2}}}^{2}+{H^4_{{-2}}}^{3} \right) {p_{{3}}}^{2}-
 \left( -3\,{H^4_{{0}}}^{2}-2\,{H^3_{{0}}}^{2}{H^3_{{-2
}}}^{2}-2\,{H^3_{{-2}}}^{2}H^4_{{0}}H^4_{{-2}}
-2\,{H^3_{{-2}}}^{4}H^4_{{0}}\right. \\  
& \left.+3\,{H^3_{{0}}}^{2}H^4_{{-2
}}+2\,H^3_{{0}}H^3_{{-2}}{H^4_{{-2}}}^{2}-8\,H^3_{
{0}}H^3_{{-2}}H^4_{{0}}+2\,H^3_{{0}}{H^3_{{-2}}}^{
3}H^4_{{-2}} \right) p_{{4}}- \left( -4\,{H^3_{{0}}}^{3}-4\,
H^3_{{-2}}{H^4_{{0}}}^{2}\right. \\ 
& \left.+6\,H^3_{{0}}H^4_{{0}}H^4_{{-2}}
-4\,H^3_{{0}}H^4_{{0}}{H^3_{{-2}}}^{2}+6
\,{H^3_{{0}}}^{2}H^3_{{-2}}H^4_{{-2}}+2\,H^4_{{0}}
H^3_{{-2}}{H^4_{{-2}}}^{2}+2\,H^4_{{0}}{H^3_{{-2}}
}^{3}H^4_{{-2}}\right. \\ 
& \left.-2\,H^3_{{0}}{H^3_{{-2}}}^{2}{H^4_{
{-2}}}^{2}-2\,H^3_{{0}}{H^4_{{-2}}}^{3} \right) p_{{3}}=0,
\end{split}
\end{equation}
where $H^3_{-2}$,$H^3_{0}$,$H^4_{-2}$ and $H^4_{0}$ obey the constraints (\ref{nlin-S3}). The (3,4) curve (\ref{(3,4)-S3}) have zero genus. 
\end{prop}
{\bf Proof } By direct calculation with the use  of polynomial form (\ref{p-S3}) of $p_j$. $\square$ \par
Comparing the results of this and previous section, one observes an essential difference between the geometrical properties of the subspaces $W_2$ and $W_3$. This is due to the quite different form of the Laurent series belonging to $W_i$ which is the consequence of a different situation with elements of the first order in $z$. Namely, though in both cases $W_i$ does not contain the element $p_1(z)$, The absence in $W_3$ of the terms of order $z$ in $p_j(z)$ leads to a strong constraints leading to the polinomiality of $p_j(z)$. \par
We note also that due to the presence of the element $p_0=1$ of zero order in $W_2$ one has $z^2 \in W_2$ while $z^2 \notin W_3$. As a consequence, for instance, one can choose $p_3$, $p_4$ and $z^2p_3$ as the generators of the algebra $A_3$ instead of $p_3$, $p_4$ and $p_5$. \par
A way to avoid the polinomiality of all $p_j(z) \in W_3$ would be to relax the condition $z^2 W_3 \subset W_3$. Since $z^2$ is not an element of $W_3$ it would be natural not to require that the product of $z^2$ and an element of $W_3$ belongs to $W_3$, but instead to require that  $z^2 W_3 \subset \Sigma_3$. The presence of the element $p_1(z)$ in $\Sigma_3$, allows us to avoid immediate constraints on $p_j(z)$ followed from the relations of the type 
(\ref{evenconst-S3}) and (\ref{oddconst-S3}). for instance, instead of the conditions (\ref{rel-S3}) one gets the following ones
\begin{equation}
 \begin{split}
  &z^2p_3 -p_5 -H^3_{-2}p_4=H^3_1p_1, \\
&z^2p_4 -p_6 -H^4_{-2}p_4=H^4_1p_1, 
 \end{split}
\end{equation}
 and so on. In virtue of the equations of this type one obtains  an infinite set of relations for $H^j_k$. Computer analysis strongly indicates that these conditions again lead to the constraint $H^j_k=0$, $k=1,2,3,\dots$, $j=3,4,5,\dots$, i.e. to the polinomiality of all $p_j(z)$.
\section{Strata $\mathbf{\Sigma_{2n}}$. Hyperelliptic curves of genus n}
Stratum $\Sigma_{2n}$ with arbitrary $n$ is characterized by $S=\{ -2n, -2n+2, -2n+4, \dots, 0, 2, 4, \dots, 2n, 2n+1, 2n+2, \dots \}$. So it does not contain, in particular, $n$ elements of the order $ 1, 3, 5, \dots, 2n-1$ and the positive order elements of the canonical basis are given by
\begin{equation}
 \label{series-S2n}
\begin{split}
 p_0=& 1+\sum_{k \geq 1} \frac{H^0_k}{z^k},\\
 p_j=& z^j+\sum_{k=0}^{j-1}H^j_{-2k-1}z^{2k+1} + \sum_{k \geq 1} \frac{H^j_k}{z^k}, \qquad j=2,4,6,\dots, 2n-2, \\
 p_j=& z^j+\sum_{k=0}^{n-1}H^j_{-2k-1}z^{2k+1} + \sum_{k \geq 1} \frac{H^j_k}{z^k} \qquad j=2n,2n+1,2n+2,2n+3,\dots\ .
\end{split}
\end{equation}
As in the previous cases the $p_j$ with negative $j$ do not should be taked into account and one has 
\begin{lem}
 A set $W_{2n}$ at fixed $H^j_k$ of Laurent series (\ref{series-S2n}) obey the condition $z^2 W_{2n} \subset W_{2n}$ and equations
\begin{equation}
\label{S2n+1-alg}
 p_j(z)p_k(z)=\sum_l C_{jk}^l p_l(z) ,\qquad j,k,l=0,2,4,\dots,2n,2n+1,2n+3,\dots
\end{equation}
\end{lem}
if and only if
\begin{equation}
 \label{Heven-S2n}
\begin{split}
 H^{2m}_k =& 0, \qquad m =0,1,2, \dots, \ k= -2n+2, -2n+4, \dots, -2, 0, 1, 2, 3, \dots, \\
H^{2m+1}_{2k} =& 0, \qquad m = 0,1,2, \dots \ k = -n , -n+1, -n+2, \dots 
\end{split}
\end{equation}
and
\begin{equation}
\label{H-S2n}
\begin{split}
&H^{2j+1}_{2(l+k)+1}-H^{2(j+k)+1}_{2l+1}-\sum_{s=-n}^{k-1}H^{2j+1}_{2s+1}H^{2(k-s)-1}_{2l+1}=0, \\
&H^{2j+1}_{2(l+k)+1}+H^{2k+1}_{2(l+j)+1}
+\sum_{s=-n}^{-1} H^{2j+1}_{2s+1}H^{2k+1}_{2(l-s)-1}
+\sum_{r=-n}^{-1} H^{2k+1}_{2r+1}H^{2j+1}_{2(l-r)-1}
+\sum_{s=0}^{l-n} H^{2j+1}_{2s+1}H^{2k+1}_{2(l-s)-1}=0.
\end{split}
\end{equation}
Rewriting equation (\ref{S2n+1-alg}) separately for $p_j$ with even and odd $j$, i.e.
\begin{equation}
\begin{split} 
 p_{2j} p_{2k} =& p_{2(j+k)}, \\
 p_{2j} p_{2k+1} =& p_{2(j+k)+1} +\sum_{s=-n}^{k-1}H^{2j+1}_{2s+1}p_{2(k-s)-1},\\
p_{2j+1} p_{2k+1} =& p_{2(j+k+1)}
+\sum_{s=-n}^j H^{2j+1}_{2s+1}p_{2(j-s)}
+\sum_{s=-n}^k H^{2k+1}_{2s+1}p_{2(k-s)} \\
&+\sum_{s=-n}^{-1}\sum_{r=-n}^{-1}H^{2j+1}_{2s+1}H^{2k+1}_{2r+1}
p_{-2(s+r+1)}+\sum_{s=-n}^{-1}\sum_{r=0}^{-s-1}H^{2j+1}_{2s+1}H^{2k+1}_{2r+1}p_{-2(s+r+1)}\\
&+\sum_{r=-n}^{-1}\sum_{s=0}^{-r-1}H^{2j+1}_{2s+1}H^{2k+1}_{2r+1}p_{-2(s+r+1)},
\end{split}
\end{equation}
one concludes that
\begin{equation}
 \begin{split}
  p_{2m}=&(z^2)^m, \\
p_{2m+1}=& \alpha(\lambda) p_{2n+1}, \qquad m=n+1,n+2,\dots, \ \lambda=z^2\\
 \end{split}
\end{equation}
for suitable $ \alpha(\lambda) \in \mathbb{C}[\lambda] $. Moreover
\begin{equation}
 \label{pp-S2n}
p_{2n+1}^2 = \lambda^{2n+1} +\sum_{k=0}^{2n} u_k \lambda^k
\end{equation}
where the coefficients $u_k$ can be obtained from
\begin{equation}
 \label{uC-S2n}
\begin{split}
 p_{2n+1}^2 = \lambda^{2n+1} +2\sum_{s=0}^{2n} H^{2n+1}_{2(n-s)+1} \lambda^s
+\sum_{k=-n}^{n+1} \sum_{s=0}^{n-k-1} H^{2n+1}_{2k+1}H^{2n+1}_{-2(s+k)-1} \lambda^s.
\end{split}
\end{equation}
Thus, one has
\begin{prop}
The stratum $\Sigma_{2n}$ for $n=2,3,4,\dots$ contains maximal subset $W_{2n}$ closed with respect to pointwise multiplication. Elements of $W_{2n}$ are vector spaces with basis given by $\langle p_i\rangle_{i=0,2,4,\dots,2n,2n+1,2n+3,\dots}$
with parameters $H^j_k$ obeying the constraints (\ref{Heven-S2n}) and (\ref{H-S2n}). 
\end{prop}
\begin{prop}
The subset $W_{2n}$ is the infinite family of infinite-dimensional commutative associative algebra $A_{2n}$ isomorphic to $\mathbb{C}[\lambda,p_{2n+1}]/ C_{2n+1}$
where $\lambda=z^2$ and
 \begin{equation}
 \label{C-S2n}
C_{2n+1}=p_{2n+1}^2 - \lambda^{2n+1} -\sum_{k=0}^n u_k \lambda^k=0
\end{equation}
and $u_k$ are given by (\ref{uC-S2n})
\end{prop}
{\bf Proof } Associativity of the algebra $A_{2n}$ is the consequence of the equivalence of the constraints (\ref{Heven-S2n}) and (\ref{H-S2n}) and the associativity conditions 
\begin{equation}
 \sum_s C_{jk}^sC_{ls}^r=  \sum_s C_{lk}^sC_{js}^r
\end{equation}
for the constants $C_{jk}^l$ defined in (\ref{S2n+1-alg}) i.e.
\begin{equation}
\begin{split}
 C_{2j,2k}^{2l}=&\delta^l_{j+k}, \\
C_{2j+1,2k}^{2l+1}=&\delta^l_{j+k} + H^{2j+1}_{2(k-l)-1},\\
C_{2j,2k}^{2l}=&\delta^l_{j+k}
+H^{2j+1}_{2(j-l)+1}
+H^{2k+1}_{2(k-l)+1} 
+\sum_{s=-n-1}^{-1}\sum_{r=-n-1}^{-1}H^{2j+1}_{2s+1}H^{2k+1}_{2r+1}\delta^l_{-2(s+r+1)}\\&
+\sum_{s=-n-1}^{-1}\sum_{r=0}^{-s-1}H^{2j+1}_{2s+1}H^{2k+1}_{2r+1}\delta^l_{-2(s+r+1)}
+\sum_{r=-n-1}^{-1}\sum_{s=0}^{-r-1}H^{2j+1}_{2s+1}H^{2k+1}_{2r+1}\delta^l_{-(s+r+1)}.  
\end{split}
\end{equation}
$\square$\par
Interpreting $\lambda, p_{2n+1}, p_{2n+3}, \dots$ as the local affine coordinates we first observe that the equation
\begin{equation}
 C_{2n+1}=0
\end{equation}
defines a hyperelliptic curve of genus $n$. It is parameterized by $2n+1$ arbitrary quantities $H^{2n+1}_{-2n+1}$, $H^{2n+3}_{-2n+1}$, $\dots$ $H^{2n+1}_{+1}$, $\dots$, $H^{2n+1}_{2n+1}$. Their variation generates an infinite family of hyperellliptic curves. \par
Each hyperelliptic curve from this family (at fixed $H^j_k$) is associated with a point of the subset $W_{2n}$. So one can refer to such points as {\it hyperelliptic points} in $\Sigma_{2n}$  and the whole $W_{2n}$ as {\it hyperelliptic subset} in $\Sigma_{2n}$.\par
One has
\begin{prop}
 The subset $W_{2n}$ in $\Sigma_{2n}$ is an infinite family of infinite-dimensional algebraic variety $\Gamma_{2n}$ defined by the relations (\ref{S2n+1-alg}),(\ref{Heven-S2n}), (\ref{H-S2n}), (\ref{C-S2n}). Its ideal is
\begin{equation}
 \label{I-S2n}
I_{2n+1}=\langle C_{2n+1}, l^{(n)}_{2n+3}, l^{(n)}_{2n+5}, \dots \rangle
\end{equation}
where $l^{(n)}_{2m+1}=p_{2m+1}-\alpha_m(\lambda) p_{2n+1}$, $m=n+1,n+2,n+3,\dots$.
\end{prop}
In other words the variety $\Gamma_{2n}$ is the intersection of the cubic $C_{2n+1}=0$ and infinite set of algebraic curves $l^{(n)}_{2m+1}$, $m=n+1,n+2,n+3,\dots$. One can easily see that the ideal $I_{2n}$ contains higher order hyperelliptic curves but all of them have genus $n$.\par
Thus stratum $\Sigma_{2n}$ is characterized by the presence of the plane hyperelliptic curves $C_{2n+1}$ of genus $n$ in every point of the closed subset $W_{2n}$. This is due to the presence of $n$ gaps (elements $p_1(z)$, $p_3(z), \dots$, $p_{2n-1}(z)$) in the basis of $W_{2n}$. The fact that for hyperelliptic curves (Riemann surfaces) of genus $n$ one has $n$ (Weierstrass) gaps in a generic point is well known in the theory of abelian functions (see e.g. \cite{Baker}). Probably  not that known observation is that these gaps and consequently the properties of corresponding algebraic curves are prescribed by the structure of the Birkhoff strata $\Sigma_{2n}$ in Gr$^{(2)}$. In different context an appearance of hyperelliptic curves in Birkhoff strata of Gr$^{(2)}$ has been observed in \cite{KK}. 
\section{Strata $\mathbf{\Sigma_{2n+1}}$}
Stratum $\Sigma_{2n+1}$, $n=2,3,4,\dots$ is characterized by $S=\{-2n-1,-2n+1,\dots,-1,1,3,\dots, 2n+1,2n+2,\dots \}$. So, the positive order elements of the canonical basis in $\Sigma_{2n+1}$ are of the form 
\begin{equation}
\label{series-S2n+1}
\begin{split}
 p_{j}(z)=&z^j+H^j_{-j+1}z^{j-2}+H^j_{j+2}z^{j-2}+\dots+H^j_0+\sum_{k \geq 1} \frac{H^j_k}{z^k}, \qquad j=1,3,\dots,2n-1 \\
p_{j}(z)=&z^j+H^j_{-2n}z^{2n}+H^j_{-2n+1}z^{2n-1}+\dots+H^j_0+\sum_{k \geq 1} \frac{H^j_k}{z^k}, \qquad j=2n+1,2n+2,\dots \ .
\end{split}
\end{equation}
As in the previous cases the $p_j$ with  $j \leq 1$ do not should be taked into account.\par
Closed subsets in $\Sigma_{2n+1}$ have different structure for different $n$. In order to see this let us begin with $\Sigma_5$. In this case the elements (\ref{series-S2n+1}) of the canonical basis are
\begin{equation}
\label{series-S5} 
\begin{split}
  p_1=&z+H^1_0+\sum_{k \geq 1} \frac{H^1_k}{z^k}, \\
p_3=&z^3+H^3_{-2}z^2+H^1_0+\sum_{k \geq 1} \frac{H^1_k}{z^k}, \\
p_j=&z^j+H^j_{-4}z^4+H^j_{-2}z^2+H^j_0+\sum_{k \geq 1} \frac{H^j_k}{z^k} \qquad j=5,6,7,\dots\ .
 \end{split}
\end{equation}
It is easy to see that the maximal closed subset $W_5$ in $\Sigma_5$ is the subset whose points are vector spaces  with  basis $(p_3,p_5,p_6,\dots)$.
\begin{lem}
 \label{lem-S5}
A set $W_5$ at fixed $H^j_k$ of the Laurent series $p_3,p_5,p_6,\dots$ obey the equations
\begin{equation}
 \label{alg-S5}
p_j(z)p_k(z) =\sum_{l=3,5,6,\dots} C_{jk} p_l(z)
\end{equation}
and the condition $z^2 W_5 \subset W_5$ if and only if $H^j_k=0$, $j=3,5,6,\dots$, $k=1,2,3,\dots$, i.e. all $p_j$
are polynomials and $H^j_k$, $k=-4,-2,0$ obey the constraints
\begin{equation}
\label{fH-S5}
 \begin{split}
& H^5_0=0 \qquad H^5_{-2}=H^3_0, \qquad H^5_{-4}=H^3_{-2}, \\
& H^6_0=-{H^3_0}^2, \qquad H^6_{-2}=-2H^3_0H^3_{-2} , \qquad H^6_{-4}=-{H^3_{-2}}^2, \\
&\dots
 \end{split}
\end{equation}
\end{lem}
The proof of the polynomiality of $p_j(z)$ is exactly the same as for $W_3$ (Lemma \ref{lem-S3}). The constraints (\ref{fH-S5}) follow from equations (\ref{alg-S5}) and the condition $z^5 W_5 \subset W_5$. For instance one has $z^2p_3=p_5$, $z^2p_5=p_7+H^5_{-4}p_6$ etc. . The constants $C_{jk}^l$ are given by
\begin{equation}
 \label{C-S5}
  C_{jk}^l=\delta_{j+k}^l+\sum_{s=0}^2 H^j_{-2s} \delta^l_{2s+k}
+\sum_{s=0}^2 H^k_{-2s} \delta^l_{2s+j}+\sum_{s,r=0}^2 H^j_{-2s} H^k_{-2r} \delta^l_{2(r+s)},  \quad j,k \geq 3
\end{equation}
where, for sake of compactess, we use $H^3_{-4}=0$. 
 As a consequence of this lemma one has 
\begin{prop}
 \label{prop-S5}
The stratum $\Sigma_5$ contains a maximal subset $W_5$ closed with respect to pointwise multiplication $W_5 \cdot W_5 \subset W_5$. Elements of $W_5$ are vector spaces $\langle p_i\rangle_{i=3,5,6,7,\dots }$ and $H^j_{-4}$, $H^j_{-2}$, $H^j_{0}$ obeying the constraints (\ref{fH-S3}).\par
Algebraically $W_5$ is an infinite family of infinite-dimensional commutative associative algebra $A_5$ of polynomials with the structure constants given by (\ref{C-S5}). Geometrically $W_5$ is the infinite algebraic variety $\Gamma_5$ defined by the equations (\ref{alg-S5}) and (\ref{fH-S3}).
\end{prop}
First equations of the set of equations (\ref{alg-S5}) are 
\begin{equation}
 \label{falg-S5}
\begin{split}
{p_3}^2=& p_6+2H^3_{-2}p_5 +2H^3_0 p_3,\\
{p_3}p_5=&p_8+2H^3_{-2}p_7+{H^3_{-2}}^2p_6+ +2H^3_0 p_5, 
\end{split}
\end{equation}
 and so on. So the algebra $A_5$ is generated by $p_3,p_5$ and $p_7$. \par
It is not also difficult to show that an ideal of the variety $\Gamma_5$ contains the family of plane (3,5) curve
\begin{equation}
 \label{(3,5)-S5}
{p_{{5}}}^{3}-{p_{{3}}}^{5}+2\,H^3_{-2}{p_{{3}}}^{3}p_{{5}}
-{H^3_{-2}}^{2}p_{{3}}{p_{{5}}}^{2}+2\,H^3_{0}{p_{{3}}}^{4}-
2\,H^3_{0}H^3_{-2}{p_{{3}}}^{2}p_{{5}}-{H^3_{0}}^{2}{p_{{3}}}^{3}=0
\end{equation}
parameterized by two variables $H^3_{-2}$ and $H^3_0$. Due to the polinomiality of $p_3$ and $p_5$ in $z$, the genus of of curve (\ref{(3,5)-S5}) is obviously equal to zero. The ideal of the varieties contains another rational plane curve given by
\begin{equation}
\label{(5,6)-S5}
\begin{split}
 & {p_{{6}}}^{5}-{p_{{5}}}^{6}+6\,H^3_{-2}p_{{5}}{p_{{6}}}^{4}
+14\,{H^3_{-2}}^{2}{p_{{5}}}^{2}{p_{{6}}}^{3}- \left( -6\,H^3_{0}
-16\,{H^3_{-2}}^{3} \right) {p_{{5}}}^{3}{p_{{6}}}^{2}
- \left( -16\,H^3_{0}H^3_{-2}-9\,{H^3_{-2}}^{4} \right) {p_{{5}}}^{4}p_{{6}} \\ &
- \left( -10\,H^3_{0}{H^3_{-2}}^{2}-2\,{H^3_{-2}}^{5} \right) {p_{{5}}}^{5}+
2\,{H^3_{0}}^{2}{p_{{6}}}^{4}+8\,{H^3_{0}}^{2}H^3_{-2}p_{{5}}{p_
{{6}}}^{3}+10\,{H^3_{-2}}^{2}{H^3_{0}}^{2}{p_{{5}}}^{2}{p_{{6}}}^{
2}\\ &
- \left( -14\,{H^3_{0}}^{3}-4\,{H^3_{-2}}^{3}{H^3_{0}}^{2}
 \right) {p_{{5}}}^{3}p_{{6}} +20\,{H^3_{0}}^{3}H^3_{-2}{p_{{5}}}^{
4}+{H^3_{0}}^{4}{p_{{6}}}^{3}+2\,{H^3_{0}}^{4}H^3_{-2}p_{{5}}{p_
{{6}}}^{2}+8\,{H^3_{0}}^{5}{p_{{5}}}^{3}=0.
\end{split}
\end{equation}
 The stratum $\Sigma_5$ exhibits the main features of higher strata $\Sigma_{4m-1}$, $(m=1,2,3,\dots)$. The maximal closed subsets $W_{4m-1}$ have the basis $(p_{2m+1}, p_{2m+3}, \dots, p_{4m-1}, p_{4m}, p_{4m+1}, \dots)$
while the stratum $\Sigma_{4m+1}$, $(m=1,2,3,\dots)$ have the basis 
$(p_{2m+1},p_{2m+3},\dots,p_{4m-1},p_{4m+1},p_{4m+2},\dots)$
with the respective $p_j$. Then one can demonstrate an analog of the Lemma \ref{lem-S5} for $\Sigma_{4m \pm 1}$
which in particular says that all $p_j(z)$ are polynomials in $z$ obeying the equations
\begin{equation}
 \label{alg-S4m-pm-1}
p_j(z)p_k(z) =\sum_{l} C_{jk}^l p_l(z), \qquad j,k,l=2m+1,2m+3,\dots
\end{equation}
together with certain constraints on $H^j_{-k}$.\par
Consequently one has closed subsets $W_{4m \pm 1}$ in $\Sigma_{4m \pm 1}$ which algebraically are commutative and associative algebras and geometrically they represent families of algebraic varieties $\Gamma_{4m \pm 1}$ defined by the equation (\ref{alg-S4m-pm-1}). Ideals of the varieties $\Gamma_{4m \pm 1}$ contain plane $(2m+1,2m+3)$ curve 
\begin{equation}
\label{(2m+1,2m+3)-S4m-pm-1}
 p_{2m+1}^{2m+3}-p_{2m+3}^{2m+1}+\dots =0, \qquad m=1,2,3,\dots
\end{equation}
of zero genus.\par
Properties of these rational curves will be discussed elsewhere.
\subsubsection*{Acknowledgements} B.K. thanks V. M. Buchstaber and B. Dubrovin  for useful discussions. This work has been partially supported by PRIN grant no 28002K9KXZ and by FAR 2009 (\emph{Sistemi dinamici Integrabili e Interazioni fra campi e particelle}) of the University of Milano Bicocca.

\end{document}